\documentclass[fleqn,usenatbib]{mnras}

\usepackage{newtxtext,newtxmath}
\usepackage[T1]{fontenc}

\let\VANthebibliography\thebibliography
\def\thebibliography{\DeclareRobustCommand{\VAN}[3]{##3}\VANthebibliography}

\usepackage{graphicx}	
\usepackage{amsmath}	
\usepackage{booktabs}	
\usepackage{subfig}   
\usepackage{multirow} 
\usepackage{xcolor}

\newcommand{\excursion}{\ensuremath{K_\nu}} 
\newcommand{\thresh}{\ensuremath{\nu} } 
\newcommand{\mach}{\ensuremath{\mathcal{M}} } 
\newcommand{\cossim}{\ensuremath{\textbf{CosSim}} } 
\newcommand{\rmsd}{\ensuremath{\textbf{RMSD}} }
\newcommand{\totalvoxels}{\ensuremath{N_\text{cells}}} 
\newcommand{\kin}{\ensuremath{\rm Kin}} 
\newcommand{\sat}{\ensuremath{\rm Sat}} 
\newcommand{\gau}{\ensuremath{\rm GRF}} 
\newcommand{\eV}{{\rm eV}}  
\newcommand{\cm}{{\rm cm}}    
\newcommand{\pc}{{\rm pc}}    
\newcommand{\kpc}{{\rm kpc}}    
\newcommand{\muG}{\mu{\rm G}} 
\newcommand{\brms}{{b_{\rm rms}}} 
\newcommand{\Fone}{\ensuremath{\phi_1}}
\newcommand{\Ftwo}{\ensuremath{\phi_2}}
\newcommand{\bbybrms}{\ensuremath{b_i / \brms}}

\def\apj{Astrophys.\ J.}

\def\mnras{Mon.\ Not.\ R.\ Astron.\ Soc.}
\def\aap{Astron.\ Astrophys.}
\def\apjl{Astrophys.\ J.\ Lett.}

\def\physrep{Phys.\ Rep.}
\def\pre{Phys.\ Rev.\ E}
\def\prl{Phys.\ Rev.\ Lett.}

\def\apjs{ApJS}

\def\pasa{PASA}

\def\nat{Nature}
\def\pasp{PASP}

\def\araa{Ann.\ Rev.\ Astron.\ Astrophys.}

\newcommand\Eq[1]{Eq.\,\ref{#1}}
\newcommand\Fig[1]{Fig.~\ref{#1}}
\newcommand\Sec[1]{Sec.~\ref{#1}}
\newcommand\Tab[1]{Table~\ref{#1}}

\newcommand\rev[1]{{#1}}
\newcommand\revb[1]{{#1}}

\title[Non-Gaussian Magnetic \rev{Structures}] {Non-Gaussian Magnetic \rev{Structures} in the Small-Scale Turbulent Dynamo}

\author[Behara \& Seta]{
	Sasi M. Behara$^{1}$\thanks{E-mail: \href{mailto:beharasasimitra211141@students.iisertirupati.ac.in}{beharasasimitra211141@students.iisertirupati.ac.in}}
	and
	Amit Seta$^{2}$\thanks{E-mail: \href{mailto:amit.seta@anu.edu.au}{amit.seta@anu.edu.au}}
	\\
	$^{1}$Department of Physics, Indian Institute of Science, Education and Research, Tirupati 517619, India  \\
	$^{2}$Research School of Astronomy and Astrophysics, Australian National University, Canberra, ACT 2611, Australia\\
}
\date{Accepted XXX. Received YYY; in original form ZZZ}
\pubyear{\the\year{}}
\begin{document}
	\label{firstpage}
	\pagerange{\pageref{firstpage}--\pageref{lastpage}}
	\maketitle
	\begin{abstract}
		The small-scale turbulent dynamo is a key mechanism for amplifying galactic magnetic fields, yet the resulting field morphology remains poorly understood. Using 3D driven turbulence simulations across a range of compressibilities, characterised by Mach number, and Minkowski functionals, we quantitatively investigate the morphology of magnetic fields generated by the small-scale turbulent dynamo in both the exponentially growing kinematic stage and the statistically steady saturated stage. In both stages and across all Mach numbers, we find that the magnetic field departs significantly from a Gaussian random field. Magnetic structures are statistically less curved and more interconnected in the saturated stage than in the kinematic stage, with these morphological differences decreasing as compressibility increases. Our work provides a quantitative description of how density fluctuations in turbulence and the back-reaction of amplified magnetic fields via the Lorentz force together shape complex, non-Gaussian magnetic structures, and offers a valuable framework for comparing simulations with \revb{both analytical models and} polarisation observations.
	\end{abstract}
	
	
	
	
	\section{Introduction} \label{sec:intro}
	
	Magnetic fields are a dynamically important component of the interstellar medium (ISM) in the Milky Way and, more broadly, in star-forming galaxies. Their average energy density ($\approx 1\,\eV\,\cm^{-3}$) is comparable to other ISM components such as thermal gas, turbulence, and cosmic rays. In particular, magnetic fields play a crucial role in star formation \citep{PattleEA2023} and in the propagation of cosmic rays \citep{RuszkowskiP2023}. However, beyond the strength of galactic magnetic fields \citep{Haverkorn2015, Beck2016, SetaMG2025}, much remains unknown, making it difficult to fully understand their role in star formation and galaxy evolution. In particular, little is known about their structural properties on smaller scales ($\lesssim 100\,\pc$), which this work aims to investigate.
	
	Magnetic fields in protogalaxies \citep[$\approx 10^{-4}\,\muG$;][]{Subramanian2016} are significantly weaker than those observed in present-day galaxies \citep[$\approx 10\,\muG$;][]{Beck2016}. This amplification of magnetic fields is attributed to a dynamo mechanism, in which the kinetic energy of turbulence is converted into magnetic energy \citep{RuzmaikinEA1988, BrandenburgS2005, Rincon2019, ShukurovS2021}. Based on the driving scale of turbulence (e.g.,~$\sim 100\,\pc$ in galaxies, considering supernova explosions as the primary drivers of the ISM turbulence), magnetic fields and dynamo processes can be categorised into large- and small-scales. The small-scale turbulent dynamo extracts turbulent kinetic energy and generates random magnetic fields on scales smaller than the driving scale of turbulence, whereas the large-scale dynamo utilises the random fields generated by the small-scale dynamo, together with differential rotation and vertical density stratification, to produce large-scale, coherent magnetic fields on $\kpc$ scales \citep[such as those observed in radio polarization observations; see][]{FletcherEA2011, Beck2015}. In this \revb{paper}, we explore the morphology of magnetic fields generated by the small-scale turbulent dynamo.

	The small-scale turbulent dynamo broadly has three stages \citep[e.g.,~see Fig.~1 in][]{SetaEA2020}. First, the \textit{kinematic} stage, where the initial weak seed field grows exponentially in strength \citep{Kazantsev1968}. Second, when the magnetic field becomes strong enough to react back on the flow via the Lorentz force, the exponential growth slows down, but the field strength continues to increase in a power-law fashion \citep[e.g.,~see Fig.~2(b) in][]{SetaF2020}. Finally, the magnetic field reaches a statistically steady state, known as the {\it saturated} stage. Both analytically \citep{ZeldovichEA1984, Subramanian1998, SchekochihinEA02} and numerically \citep{SchekochihinEA2004, HaugenEA2004, seta_saturation_2021, SurS2024}, the magnetic fields generated by the small-scale turbulent dynamo are found to be non-Gaussian across all three stages. \revb{Fundamentally, such non-Gaussianity is essential for the nonlinear energy cascade in hydrodynamic and magnetohydrodynamic turbulence \citep{SheLeveque1994, MullerBiskamp2000, LiMeneveau2005}. For the small-scale turbulent dynamo}, in the kinematic stage using simple, random flows \citep{WilkinBS2007}, and in both the kinematic and saturated stages of numerically driven subsonic turbulent flows \citep{SetaEA2020}, the magnetic fields are shown to exhibit more filamentary structures. 
	
	In this work, we use Minkowski functionals (MFs) to describe and quantify the level of non-Gaussianity and characterise 3D structures in magnetic fields generated by the small-scale turbulent dynamo. MFs were originally developed to study the large-scale cosmological structures \citep{Mecke1994} and have since been developed into powerful statistical tools \citep{varun_sahini_shapefinders_1998}.  Usually, the two-point statistics, power spectra \citep[or equivalently correlation or structure functions, e.g.,~see][for a discussion]{SetaEA2023, MethaB2025}, are used to study random fields in astrophysics, especially for turbulent phenomena. They fail to capture the higher orders of correlations, which means it is possible to construct two random fields with the same power spectrum but very different morphological features and levels of non-Gaussianity \citep[e.g.,~see Fig.~1 in][for a demonstration]{SetaEA2018}. MFs are better tools for studying random structures as they provide statistically unbiased descriptors which contain features of $n$-point correlation functions, for any order $n$ \citep{Mecke1994}. 
	
	It is important to characterise 3D magnetic structures for a variety of reasons, most importantly to understand the physics of small-scale turbulent dynamo \citep[][also, our aim in this work]{SetaEA2020}, decipher the acceleration and diffusion of cosmic rays \citep{ShukurovEA2017, Lemoine2023, Kempski_2023, Butsky_2024, ReichherzerEA2025, LubkeEA2025}, and in the interpretation of projected 2D polarisation observations \citep{GaenslarEA2011, ZaroubiEA2015, ErcegEA2022}. Most such and related previous work characterise only the strong field or low-volume filling magnetic structures using probability distribution functions (one-point statistics) or power spectra (two-point statistics). Even those that characterise morphology \citep[e.g.,][]{WilkinBS2007, Zhdankin_2013, SetaEA2020, DwivediEA2024, DuttaEA2024} focus on low-volume, strong field regions and only use the scalar quantities, such as the strength of the magnetic field or polarised emissivity \revb{and primarily explore low volume filling, strong field/emissivity regions}. We, for the first time, apply MFs to the full magnetic field vector to characterise the 3D morphology of magnetic structures. \revb{This approach reveals the global topology,  interconnectedness, and curvature of the field, independent of field strength. By analysing all vector values, we capture the changes induced by saturation in the weak and intermediate fields, which dominate the volume, but were beyond the scope of previous studies.}  
	
	\rev{We study the morphology of non-Gaussian structures} as a function of the dynamo stage and the compressibility of the turbulence.  These different Mach numbers are representative of the different ISM phases \citep[overall and approximately, hot, ionised ISM is subsonic, warm, atomic ISM is transsonic, and cold, molecular ISM is supersonic; see][]{Ferriere2020, SetaF2022}. Such an analysis helps us to characterise non-Gaussian magnetic fields,  better understand the saturation mechanism of the small-scale turbulent dynamo, and systematically compare simulations \revb{with both analytical models and} observations.
	
	The paper is organised as follows. We describe our numerical simulations and the method used to extract the Minkowski functionals in \Sec{sec:methods}. In \Sec{sec:res}, we present our results, examine their dependence on the stage of the small-scale turbulent dynamo and the Mach number, and quantify deviations from Gaussianity. \revb{In \Sec{sec:dis}}, we discuss our results in the context of small-scale turbulent dynamo saturation. Finally, we conclude in \Sec{sec:con}.
	
	\section{Methods} \label{sec:methods}
	\subsection{Simulations}
	\label{sec:simulations}
	We utilise simulation data from the work of \cite{seta_saturation_2021}, in which the authors investigated the saturation mechanism of the small-scale turbulent dynamo at different Mach numbers. The simulations were performed using a modified version of the \texttt{FLASH} code \citep[v4;][]{WaaganFK2011}, solving the non-ideal, compressible MHD equations for an isothermal gas. Explicit, constant viscosity and resistivity were employed.
	
	The simulations were carried out in a 3D Cartesian numerical domain with triply-periodic boundary conditions, discretised on a uniform grid with $512^3$ points. Each simulation was initialised with uniform density, zero velocity, and a very weak random seed magnetic field (plasma beta of $2.5 \times 10^{13}$). Turbulence was driven numerically using continuous, solenoidal forcing at half the domain size via an Ornstein-Uhlenbeck process \citep{FederrathEA2010}. Both the hydrodynamic and magnetic Reynolds numbers are fixed at $2000$ \rev{, which is well resolved for these choices of parameters  \citep[see Fig.~8 in][]{shivakumar_and_federrath_2025}.} \revb{This also implies that the Prandtl number, which is the ratio of magnetic to hydrodynamic Reynolds number, is $1$.} The key parameter varied across simulations was the turbulent Mach number, \mach, which is the ratio of the turbulent velocity to the sound speed and characterises the compressibility in the medium. $\mach$ was varied from $0.1$ (subsonic) to $10$ (highly supersonic). For further details on the simulation setup and parameters, refer to \cite{seta_saturation_2021}.
	
	For all $\mach$, the simulations exhibit the small-scale turbulent dynamo: an exponential growth phase during the kinematic stage, followed by a slower, power-law increase in the magnetic field strength, and finally, a statistically steady state in the saturated stage. For all $\mach$, in both the kinematic and saturated stages, 1D probability distribution functions show that the magnetic fields are non-Gaussian \citep[see Fig.~7 and Table 1 in][]{seta_saturation_2021}. 
	
\begin{figure*}
		\centering
		\includegraphics[width=\textwidth]{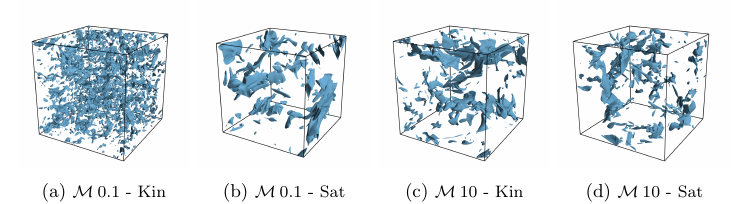}
		\caption{Excursion sets of the $x$-component of the magnetic field for $b_x/\brms > 1.2$ for $\mach=0.1$ (a, b) and $10$ (c, d) simulations in their corresponding kinematic ($\kin$; a, c) and saturated ($\sat$; b, d) stages. The magnetic structures are significantly different between the $\kin$ and $\sat$ cases for $\mach=0.1$, and such differences are less pronounced for $\mach=10$.}
		\label{fig:excursion_set_visualization}
	\end{figure*}

	\Fig{fig:excursion_set_visualization} shows the excursion set for the dynamo-generated magnetic fields for $\mach=0.1$ and $10$ in their kinematic ($\kin$) and saturated ($\sat$) stages. Visually, the structures show different features, especially when comparing the $\kin$ and $\sat$ cases for $\mach=0.1$. In this work, we quantify such differences to explore the magnetic field morphology and study deviations from Gaussianity using Minkowski functionals. 
	
	\subsection{Minkowski Functionals (MFs)}
	\label{sec:minkowski_functionals}
	
	To move beyond two-point statistics such as power spectrum and correlation functions and characterise the full morphology of the magnetic fields, we use Minkowski functionals (MFs). The Minkowski functionals are a set of morphological descriptors sensitive to the field's geometry and topology. For a 3D scalar field, $\phi(\textbf{x})$, we construct an excursion set, $K_\thresh \left( \phi \right) = \{ \textbf{x} | \phi(\textbf{x})>\thresh \}$, which is the set of all points above a given threshold, $\thresh$. The four MFs of the excursion set are \citep{schmalzing_minkowski_1995}
	\begin{enumerate}
		\item $V_0$: the volume fraction of the set
		\item $V_1$: the surface area of the set's boundary
		\item $V_2$: the integral of the mean curvature of the surface
		\item $V_3$: the Euler characteristic, quantifying the excursion 
		set's net connectivity.
	\end{enumerate}
	
	\begin{table}
		\centering
		\caption{Definition of the terms obtained by counting and used in \Eq{eq:crofton_discrete_1} -- \ref{eq:crofton_discrete_4} for computing MFs.}
		\begin{tabular}{|c|l|}
			\hline
			Quantity & Definition \\
			\hline
			$N_0(\excursion)$ & Total number of vertices (points) in $\excursion$ \\
			$N_1(\excursion)$ & Total number of edges in $\excursion$ \\
			$N_2(\excursion)$ & Total number of faces (surfaces) in $\excursion$ \\
			$N_3(\excursion)$ & Total number of cubes (volumetric cells) in $\excursion$ \\
			$\totalvoxels$ & Total number of cells in the grid (here, $512^3$) \\
			\hline
		\end{tabular}
		\label{tab:counts}
	\end{table}
	
	The MFs for a field on a discrete grid can be computed by counting elementary lattice components \citep{schmalzing_beyond_1997}. The MFs of a 3D structure mapped onto a discrete grid are given by  
	\begin{align} 
		& V_0(\excursion) = {\totalvoxels}^{-1}\, N_3(\excursion), \label{eq:crofton_discrete_1} \\ 
		& V_1(\excursion) = {\totalvoxels}^{-1} \left(-\dfrac{2}{3}\, N_3(\excursion) + \dfrac{2}{9}\, N_2(\excursion) \right), \label{eq:crofton_discrete_2} \\
		& V_2(\excursion) = {\totalvoxels}^{-1} \left(\dfrac{2}{3}\, N_3(\excursion) - \dfrac{4}{9}\, N_2(\excursion) + \dfrac{2}{9}\, N_1(\excursion) \right), 
		\label{eq:crofton_discrete_3} \\
		& V_3(\excursion) = {\totalvoxels}^{-1} \left(-N_3(\excursion) + N_2(\excursion) - N_1(\excursion) + N_0(\excursion) \right), \label{eq:crofton_discrete_4}
	\end{align}
	where Table \ref{tab:counts} defines $N_0, N_1, N_2, N_3,$ and $\totalvoxels$. The values obtained from \Eq{eq:crofton_discrete_1} -- \ref{eq:crofton_discrete_4} are inherently dimensionless and normalised by the simulation volume, allowing for direct comparison across simulations.
	
	For each magnetic field component ($b_x, b_y, b_z$) and for every $\mach$ in the $\kin$ and $\sat$ stages of the small-scale turbulent dynamo, we compute the  MFs at a given normalised threshold ($\thresh$ in \Eq{eq:crofton_discrete_1} -- \ref{eq:crofton_discrete_4}), $b_x/\brms, b_y/\brms, b_z/\brms$ ($\brms$ refers to the root-mean-square of the magnetic fields over the entire domain). This threshold is chosen to be in the range $-10$ to $10$ with $500$ equispaced points. \rev{By computing the MFs across the entire possible range of threshold, we are able to not just characterise the low-volume filling or strong field structures, but all structures that fill up the volume, thereby enabling us to characterise the entire field.} Since the turbulence driving and dynamo-generated magnetic fields are isotropic, these MFs are then averaged across the three components as a function of their normalised threshold. This normalised threshold is then represented as $\bbybrms$. Next, we calculate the mean and standard deviation of the MFs over $10$ independent realisations for each case to capture the effects of random realisations in turbulence.
	
	\subsection{Constructing Gaussian Random Fields from the Non-Gaussian Dynamo-Generated fields}
	
	A Gaussian random field (GRF) is completely determined by its two-point statistics, such as the power spectrum. In order to construct a GRF with the same power spectrum as the dynamo-generated field, we use the following process \citep{ShukurovEA2017}, 
	\begin{equation}
		\phi_{\gau} = \mathcal{F}^{-1}\left( \mathcal{F}(\phi) e^{i \theta} \right),
	\end{equation}
	where $\mathcal{F}$ is the Fourier transform operation, $\theta$ is a random phase sampled uniformly at each point between $0$ and $2\pi$, $\phi$ represents the dynamo-generated non-Gaussian random field, and  $\phi_{\gau}$ represents the corresponding Gaussianised version with exactly the same power spectrum. These are represented by $\gau$ in the simulation label. 
	
	\subsection{Quantification Metrics}
	\label{sec:analysis_metrics}
	To quantitatively compare the MFs of two different fields (whether between Gaussian vs. non-Gaussian or $\kin$ vs. $\sat$ cases), we introduce two comparison metrics, \cossim and \rmsd. 
	
	Let the sets $\{p_\thresh\}$ and $\{q_\thresh\}$ be any one of the MFs of any two fields, \Fone and \Ftwo, evaluated at a range of thresholds \thresh. The cosine similarity is defined as
	\begin{equation}
		\cossim({\rm \Fone}, {\rm \Ftwo}) =\frac{\sum_\thresh p_\thresh q_\thresh }{\sqrt{\sum_\thresh p_\thresh^2 } \sqrt{\sum_\thresh q_\thresh^2}}.
		\label{eq:cosine_similarity}
	\end{equation}
	This measures how similar two MFs are and it takes values between $-1$ and $1$, where $1$ represents a perfect match between the shapes of the MF vs. \thresh profiles for the two fields and deviations from $1$ quantify differences in those profiles. 
	
	We further quantify the level of deviation between the two profiles using the root mean square deviation, $\rmsd$, defined as
	\begin{equation}
		\rmsd({\rm \Fone}, {\rm \Ftwo}) = \left({n_\thresh}^{-1} \sum_\thresh (p_\thresh - q_\thresh)^2 \right)^{1/2},
		\label{eq:rmsd_similarity}
	\end{equation}
	where $n_\thresh$ represents the number of $\thresh$ values (here, $n_\thresh = 500$ in the range $-10$ to $10$). \rmsd is another measure of how similar two MF profiles are and it takes values between $0$ and $\infty$, where $0$ refers to a perfect match.
	
	We note that \cossim is sensitive only to the shape of the profiles, while \rmsd is sensitive to both the shape and amplitude of the profiles. A high \cossim ($\approx 1$) would indicate that the $\rmsd$ value is entirely capturing the amplitude differences between the profiles. However, when \cossim is not close to $1$, the \rmsd value gives a combined result of both the differences in the shape and the amplitude, making the value hard to interpret. Given this, we interpret $\rmsd$ only when the value of $\cossim$ is close to 1 but for completeness, we report both the $\cossim$ and $\rmsd$ for all the cases.
	
	\section{Results} \label{sec:res}
	\begin{figure*}
		\centering
		\includegraphics[width=\textwidth]{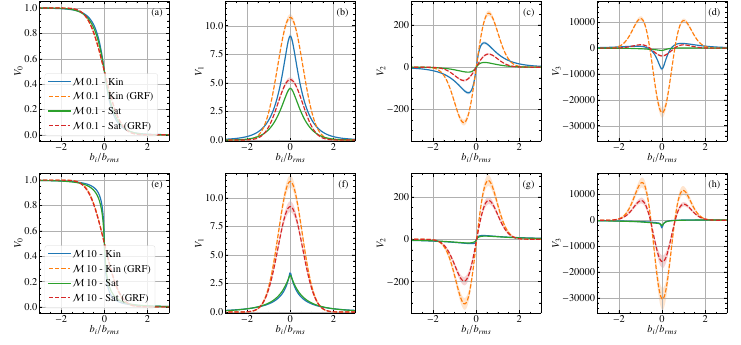}
		\caption{Minkowski functionals ($V_0, V_1, V_2,$ and $V_3$) as a function of the threshold, $\bbybrms$, for simulated magnetic fields (solid lines) and their Gaussianised versions with the same power spectra (\gau, dashed lines) in the kinematic ($\kin$) and saturated ($\sat$) stages for $\mach = 0.1$ (a-d) and $\mach = 10$ (e-h) simulations.}
		\label{fig:MFs}
	\end{figure*}

	\begin{table*}
		\centering
		\caption{Quantitative comparisons of differences in profiles of MFs vs. $\bbybrms$ (\Fig{fig:MFs}). \cossim (\Eq{eq:cosine_similarity}) compares the similarity of non-Gaussian and Gaussian (GRF) profiles with the same spectra for kinematic (\kin) vs. saturated (\sat) stages and also between the $\kin$ and $\sat$ stages. $\cossim=1$ implies the profiles are the same in shape. \rmsd (\Eq{eq:rmsd_similarity}) quantifies the difference between pairs of profiles. We interpret $\rmsd$ only when the value of $\cossim$ is close to 1 but for completeness, we report both the $\cossim$ and $\rmsd$ for all the cases.}
		\label{tab:quant}
		\resizebox{\textwidth}{!}{
			\begin{tabular}{llllllll}
				\hline
				\textbf{Functional}                            & \textbf{\mach}            & $\mathbf{\cossim (\kin, \gau)}$  & $\mathbf{\rmsd (\kin, \gau)}$    & $\mathbf{\cossim (\sat, \gau)}$  & $\mathbf{\rmsd (\sat, \gau)}$    & $\mathbf{\cossim (\kin, \sat)}$   & $\mathbf{\rmsd (\kin, \sat)}$    \\ \hline
				\multicolumn{1}{|l|}{\multirow{4}{*}{$V_{0}$}} & \multicolumn{1}{l|}{0.1}  & \multicolumn{1}{l|}{1.00 ± 0.01} & \multicolumn{1}{l|}{0.02 ± 0.01} & \multicolumn{1}{l|}{1.00 ± 0.01} & \multicolumn{1}{l|}{0.01 ± 0.01} & \multicolumn{1}{l|}{1.00 ± 0.01}  & \multicolumn{1}{l|}{0.01 ± 0.01} \\
				\multicolumn{1}{|l|}{}                         & \multicolumn{1}{l|}{2.0}  & \multicolumn{1}{l|}{1.00 ± 0.01} & \multicolumn{1}{l|}{0.03 ± 0.01} & \multicolumn{1}{l|}{1.00 ± 0.01} & \multicolumn{1}{l|}{0.02 ± 0.01} & \multicolumn{1}{l|}{1.00 ± 0.01}  & \multicolumn{1}{l|}{0.01 ± 0.01} \\
				\multicolumn{1}{|l|}{}                         & \multicolumn{1}{l|}{5.0}  & \multicolumn{1}{l|}{1.00 ± 0.01} & \multicolumn{1}{l|}{0.04 ± 0.01} & \multicolumn{1}{l|}{1.00 ± 0.01} & \multicolumn{1}{l|}{0.03 ± 0.01} & \multicolumn{1}{l|}{1.00 ± 0.01}  & \multicolumn{1}{l|}{0.01 ± 0.01} \\
				\multicolumn{1}{|l|}{}                         & \multicolumn{1}{l|}{10.0} & \multicolumn{1}{l|}{1.00 ± 0.01} & \multicolumn{1}{l|}{0.04 ± 0.01} & \multicolumn{1}{l|}{1.00 ± 0.01} & \multicolumn{1}{l|}{0.04 ± 0.01} & \multicolumn{1}{l|}{1.00 ± 0.01}  & \multicolumn{1}{l|}{0.01 ± 0.01} \\ \hline
				\multicolumn{1}{|l|}{\multirow{4}{*}{$V_{1}$}} & \multicolumn{1}{l|}{0.1}  & \multicolumn{1}{l|}{0.99 ± 0.01} & \multicolumn{1}{l|}{0.73 ± 0.01} & \multicolumn{1}{l|}{1.00 ± 0.01} & \multicolumn{1}{l|}{0.30 ± 0.01} & \multicolumn{1}{l|}{1.00 ± 0.01}  & \multicolumn{1}{l|}{0.85 ± 0.01} \\
				\multicolumn{1}{|l|}{}                         & \multicolumn{1}{l|}{2.0}  & \multicolumn{1}{l|}{0.98 ± 0.01} & \multicolumn{1}{l|}{1.56 ± 0.05} & \multicolumn{1}{l|}{0.99 ± 0.01} & \multicolumn{1}{l|}{0.61 ± 0.01} & \multicolumn{1}{l|}{1.00 ± 0.01}  & \multicolumn{1}{l|}{0.13 ± 0.01} \\
				\multicolumn{1}{|l|}{}                         & \multicolumn{1}{l|}{5.0}  & \multicolumn{1}{l|}{0.97 ± 0.01} & \multicolumn{1}{l|}{2.34 ± 0.04} & \multicolumn{1}{l|}{0.98 ± 0.01} & \multicolumn{1}{l|}{1.21 ± 0.01} & \multicolumn{1}{l|}{1.00 ± 0.01}  & \multicolumn{1}{l|}{0.05 ± 0.01} \\
				\multicolumn{1}{|l|}{}                         & \multicolumn{1}{l|}{10.0} & \multicolumn{1}{l|}{0.96 ± 0.01} & \multicolumn{1}{l|}{1.98 ± 0.02} & \multicolumn{1}{l|}{0.97 ± 0.01} & \multicolumn{1}{l|}{1.49 ± 0.02} & \multicolumn{1}{l|}{1.00 ± 0.01}  & \multicolumn{1}{l|}{0.05 ± 0.01} \\ \hline
				\multicolumn{1}{|l|}{\multirow{4}{*}{$V_{2}$}} & \multicolumn{1}{l|}{0.1}  & \multicolumn{1}{l|}{0.96 ± 0.01} & \multicolumn{1}{l|}{38.1 ± 0.6}  & \multicolumn{1}{l|}{0.98 ± 0.01} & \multicolumn{1}{l|}{10.5 ± 0.2}  & \multicolumn{1}{l|}{1.00 ± 0.01}  & \multicolumn{1}{l|}{26.2 ± 0.3}  \\
				\multicolumn{1}{|l|}{}                         & \multicolumn{1}{l|}{2.0}  & \multicolumn{1}{l|}{0.87 ± 0.01} & \multicolumn{1}{l|}{64.6 ± 3.6}  & \multicolumn{1}{l|}{0.96 ± 0.01} & \multicolumn{1}{l|}{16.8 ± 0.2}  & \multicolumn{1}{l|}{0.98 ± 0.01}  & \multicolumn{1}{l|}{7.33 ± 0.13} \\
				\multicolumn{1}{|l|}{}                         & \multicolumn{1}{l|}{5.0}  & \multicolumn{1}{l|}{0.82 ± 0.01} & \multicolumn{1}{l|}{97.3 ± 3.5}  & \multicolumn{1}{l|}{0.89 ± 0.01} & \multicolumn{1}{l|}{34.3 ± 0.2}  & \multicolumn{1}{l|}{0.98 ± 0.01}  & \multicolumn{1}{l|}{4.45 ± 0.12} \\
				\multicolumn{1}{|l|}{}                         & \multicolumn{1}{l|}{10.0} & \multicolumn{1}{l|}{0.83 ± 0.01} & \multicolumn{1}{l|}{69.6 ± 1.1}  & \multicolumn{1}{l|}{0.86 ± 0.01} & \multicolumn{1}{l|}{45.1 ± 0.7}  & \multicolumn{1}{l|}{0.99 ± 0.01}  & \multicolumn{1}{l|}{1.27 ± 0.05} \\ \hline
				\multicolumn{1}{|l|}{\multirow{4}{*}{$V_{3}$}} & \multicolumn{1}{l|}{0.1}  & \multicolumn{1}{l|}{0.90 ± 0.01} & \multicolumn{1}{l|}{3917 ± 63}   & \multicolumn{1}{l|}{0.65 ± 0.02} & \multicolumn{1}{l|}{496 ± 11}    & \multicolumn{1}{l|}{0.80 ± 0.01}  & \multicolumn{1}{l|}{1000 ± 22}   \\
				\multicolumn{1}{|l|}{}                         & \multicolumn{1}{l|}{2.0}  & \multicolumn{1}{l|}{0.50 ± 0.03} & \multicolumn{1}{l|}{4881 ± 374}  & \multicolumn{1}{l|}{0.50 ± 0.02} & \multicolumn{1}{l|}{875 ± 16}    & \multicolumn{1}{l|}{-0.39 ± 0.02} & \multicolumn{1}{l|}{1003 ± 26}   \\
				\multicolumn{1}{|l|}{}                         & \multicolumn{1}{l|}{5.0}  & \multicolumn{1}{l|}{0.33 ± 0.04} & \multicolumn{1}{l|}{8518 ± 365}  & \multicolumn{1}{l|}{0.30 ± 0.02} & \multicolumn{1}{l|}{2084 ± 19}   & \multicolumn{1}{l|}{-0.51 ± 0.04} & \multicolumn{1}{l|}{818 ± 27}    \\
				\multicolumn{1}{|l|}{}                         & \multicolumn{1}{l|}{10.0} & \multicolumn{1}{l|}{0.48 ± 0.02} & \multicolumn{1}{l|}{5647 ± 139}  & \multicolumn{1}{l|}{0.42 ± 0.02} & \multicolumn{1}{l|}{2999 ± 77}   & \multicolumn{1}{l|}{0.97 ± 0.01}  & \multicolumn{1}{l|}{70.4 ± 15.2} \\ \hline
			\end{tabular}
		}
	\end{table*}
	
	\begin{figure*}
		\centering
		\includegraphics[width=\textwidth]{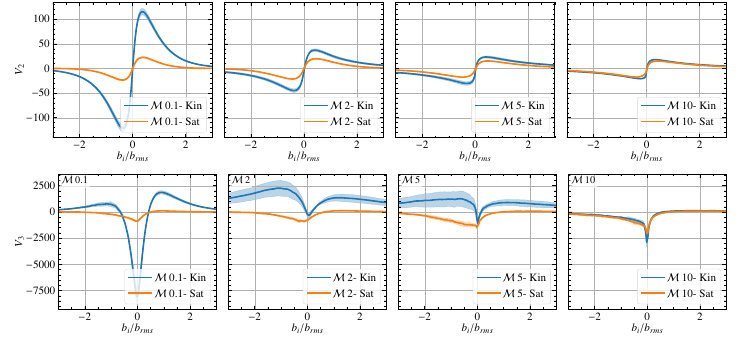}
		\caption{Comparison of $V_2$ (a-d; quantifying the mean curvature) and $V_3$ (e-h; quantifying the connectedness of the structures) for the $\kin$ and $\sat$ stages at $\mach = 0.1$ (a, e), $2$ (b, f), $5$ (c, g), and $10$ (d, h), which further highlights the morphological changes due to the turbulent dynamo saturation across Mach numbers.}
		\label{fig:MFs_2_3}
	\end{figure*}
	
	\subsection{Morphological Signatures of Gaussian Random Fields} \label{sec:GRF}
	For a \gau, the properties of MFs as a function of $\thresh$ are well-known and derived analytically \citep{Tomita1993, schmalzing_beyond_1997}. For all $\mach$ and dynamo stages, the analytical results in the shape of the profiles agree well with the numerically obtained ones for the Gaussianised version of dynamo-generated fields (the dotted lines in \Fig{fig:MFs}). For these Gaussianised fields, the shape of numerically evaluated MFs agrees well with the analytical expectations ($\cossim \approx 1$) but their amplitudes show some differences ($\rmsd \ne 0$). This is a result of various numerical approximations, especially the grid discretisation, and is well established in the literature \citep[e.g.,~see Fig.~2 in][]{pranav_topology_2019}. Thus, for our analysis, we always compare the numerically evaluated MFs between the non-Gaussian fields and their corresponding Gaussianised versions.
	
	For an easier comparison with the non-Gaussian fields, we describe the properties of MFs vs. $\thresh = \bbybrms$ profile for a GRF, shown via the dotted lines in \Fig{fig:MFs}, below.
	\begin{itemize}
		\item $V_0$ is a sigmoid-like curve, which is in essence a reversed cumulative distribution function of the field values. 
		\item $V_1$ is a bell-shaped curve centred at 0, indicating that the excursion set with the mean value as the $\bbybrms$ has the largest surface area. As $\bbybrms$ moves away from the mean, we select smaller and more isolated hot (representing values higher than the mean) or cold (representing values lower than the mean) spots, both equally likely, as implied by the symmetry of the curve.
		\item $V_2$ is an anti-symmetric curve; however, the sign is just an indication of concave or convex structures. Positive values at positive $\bbybrms$ correspond to outward-curving isolated structures (hot spots), while negative values at negative thresholds indicate inward-curving voids (cold spots)
		\item $V_3$ gives us a measure of how interconnected the field is at a given threshold, $\bbybrms$.  For a GRF, it is symmetric around zero, with positive peaks at extreme thresholds (isolated hot or cold spots) demonstrating disconnected structures at higher values and a highly negative trough at the mean, demonstrating highly connected structures. \revb{A high negative value signifies an interconnected, complex web-like structure with many tunnels and handles, which is described as a `sponge-like' configuration topologically \citep{gott_sponge-like_1986}\footnote{\revb{We note that the use of the term ‘sponge-like’ in this paper refers exclusively to a topological property, i.e., it is independent of length scales and concerns only connectivity. This usage is standard in topological studies of the large-scale structure of the Universe \citep[e.g., see Sec.~3.3.2 in][]{pranav_topology_2019}. We also emphasise that geometric properties, such as curvature ($V_2$), provide complementary but distinct information.}}}
	\end{itemize}
	
	\subsection{Morphological Signatures of the Non-Gaussian Dynamo-Generated Magnetic Fields} \label{sec:NGRF}
	
	\subsubsection{Gaussian vs. Non-Gaussian Fields} \label{sec:GRFNGRF}
	From  \Fig{fig:MFs}, \Fig{fig:MFs_2_3}, and \Tab{tab:quant}, we observe that the morphology of the dynamo magnetic field deviates significantly from \gau\, with the same power spectrum, and these deviations change with both the dynamo stage and $\mach$ of the turbulence driving.
	
	As shown in \Fig{fig:MFs}, the MFs for the dynamo\rev{-generated} magnetic fields and corresponding \gau~at \mach of 0.1 and 10, the differences in the MF profiles are evident in their amplitude for $V_1$ and $V_2$ ($\cossim \approx 1$) and also the profile shape for $V_3$ ($\cossim \ne 1$). $V_0$, the volume fraction, offers less sensitive discrimination of the morphological features for all the cases ($\cossim \approx 1$ and $\rmsd \approx 0$). For $V_1$ (surface area), when comparing both the $\kin$ and $\sat$ stages with their respective \gau s, the profiles are similar in shape ($\cossim \approx 1$) but the amplitude shows slight differences ($\rmsd > 0$). At $\bbybrms$ around $0$, the \gau~fields show higher $V_1$, representing that those fields occupy a larger surface area. At higher $|b/\brms|$ ($\gtrsim 2.5$), the non-Gaussian fields occupy non-zero surface area and $\gau$\,values $\approx 0$, so, \rev{GRFs} show $V_1 \approx 0$ \rev{at those thresholds}. $V_2$ (mean curvature) shows the most notable difference in the amplitude between the non-Gaussian and Gaussian fields. The \gau s always show significantly higher curvature than the non-Gaussian fields for all $\bbybrms$. This fundamentally demonstrates that the turbulent dynamo mechanism inherently generates fields with much lower curvature than the \gau, \rev{the non-Gaussian fields have more elongated structures.} Finally, $V_3$ (measure of interconnectedness in structures) shows a notable distinction in the shape when compared to \gau s ($\cossim < 1$). In \gau s, we see positive peaks for positive and negative values at $|b/\brms| \approx 1$, which are indicative of isolated and disconnected hot or cold spots. On the other hand, the dynamo-generated fields are consistently negative or lower across the entire range of $\bbybrms$. This suggests that the magnetic structures generated by the turbulent dynamo maintain a more interconnected, sponge-like topology, which is a fundamental departure from $\gau$, even when the power spectrum and thus, by extension, the magnetic correlation length scales are the same for both the non-Gaussian and Gaussian fields. From Table \ref{tab:quant}, we observe that these trends and thus the level of non-Gaussianity increase with increasing \mach till $\mach=5$ and then are similar/decrease for $\mach=10$. This quantitatively demonstrates that the dynamo-generated fields are inherently non-Gaussian, and their deviation from Gaussianity becomes more pronounced as compressibility increases. 
	
	\subsubsection{Kinematic vs. Saturated Stages} \label{sec:kinsat}
	\Fig{fig:MFs_2_3} show the MFs, $V_2$ (a-d) and $V_3$ (e-h), in both the $\kin$ and $\sat$ stages for all the $\mach$ numbers. $V_2$ is very similar between the $\kin$ and $\sat$ stages, i.e., $\cossim (\kin, \sat) \approx 1$, but the $\kin$ stage shows significantly higher amplitude with $\rmsd(\kin, \sat) > 0$ (see last column in \Tab{tab:quant}). Thus, on average, the magnetic structures in the $\kin$ stage are more curved than those in the $\sat$ stage.  This \revb{is consistent with} the result of \citealt{seta_saturation_2021}, where computing the magnetic correlation length, they find that the process of saturation leads to the formation of more coherent structures.
	
	For $V_3$, both the shape and the amplitude differ between the $\kin$ and $\sat$ profiles but overall, the $\sat$ case shows more negative values than the $\kin$ case. Thus, there are relatively more disconnected hot/cold spots in the kinematic stages and the magnetic field approaches a more interconnected, sponge-like topology as the turbulent dynamo saturates.
	
	For both $V_2$ and $V_3$, we note that the differences between the $\kin$ and $\sat$ profiles decrease as the \rev{mach number increases}, suggesting that the compressibility plays a significant role in the morphology of the dynamo-generated magnetic fields and it reduces the effect of the magnetic field's back-reaction.
	
	\section{Discussion} \label{sec:dis}
	
	\subsection{Insights into the Turbulent Dynamo Saturation Mechanism} \label{sec:disa}
	
	The evolution of magnetic field morphology between the $\kin$ and $\sat$ stages reveals a strong dependence on $\mach$, especially via the amplitude change in $V_2$ profiles and also the shape change in $V_3$ profiles (see \Fig{fig:MFs_2_3} and \Sec{sec:kinsat}). This change in behaviour indicates that the impact of the saturation mechanism is fundamentally different in manner at different Mach numbers, being divided into three (subsonic, moderately supersonic, and highly supersonic) regimes. We explain these trends through the interplay between the density fluctuations introduced by the turbulence and the magnetic field's back reaction via the Lorentz force for the turbulent dynamo 
	
	In the subsonic regime ($\mach=0.1$), the turbulence is largely incompressible, giving rise to negligible density fluctuations, and thus it has a relatively mild impact on the field structure. The non-Gaussianity and morphology of magnetic structures are primarily driven by the turbulent dynamo mechanism, and on saturation, the Lorentz force organises the field into comparatively less curved, more interconnected magnetic structures. 
	
	In the moderately supersonic regime ($\mach = 2, 5$), the significant density fluctuations dominate the morphology of the field in the kinematic stage, leading to a highly fragmented topology with several isolated structures. The Lorentz force then \rev{extends} these isolated structures into more interconnected, sponge-like magnetic structures. This is further demonstrated by a negative $\cossim (\kin, \sat)$ for $V_3$ (see \Tab{tab:quant}), which presents an anti-correlation. \rev
	
	In the highly supersonic regime ($\mach = 10$), the strong shocks create a complex, interconnected, and highly non-Gaussian morphology from the outset, even in the kinematic stage. This initial morphology is almost `locked in' early in the evolution. The saturation proceeds without significant changes to the topology, as demonstrated by $\cossim \approx 1$ and lower $\rmsd$ for $V_3$ than other $\mach$ cases and also relatively lower $\rmsd$ for other MFs.
	
	\rev{
		The saturation mechanism of the small-scale turbulent dynamo might be regime-dependent. The simulations used in this work probe the viscous-resistive regime, where saturation is governed by the back-reaction of the Lorentz force. In deeper detail, this back-reaction manifests through local processes, such as stretching, compression, advection, and diffusion of magnetic field lines \citep{seta_saturation_2021, SurS2024}. These are usually determined based on point-to-point computations \citep[e.g.,~see Eq.~8 in][]{SurS2024} and to connect our results to these processes requires computations on a structural basis (global analysis), which we aim to do in the future. Furthermore, at very high Reynolds numbers (magnetic Reynolds number $\gtrsim 10^4$ and hydrodynamic Reynolds number $\gtrsim 10^3$), the fields might be unstable to fast magnetic reconnection \citep{Galishnikova2022, Schekochihin2022}, which might further change the morphology of the field. However, we emphasise that our method is model-agnostic and can be applied to structures in a wide range of magnetohydrodynamic simulations.
	}
	
	\subsection{\revb{Connections with Analytical Paradigms} \label{sec:disb}}
	\begin{figure}
		\centering
		\includegraphics[width=1.05\columnwidth]{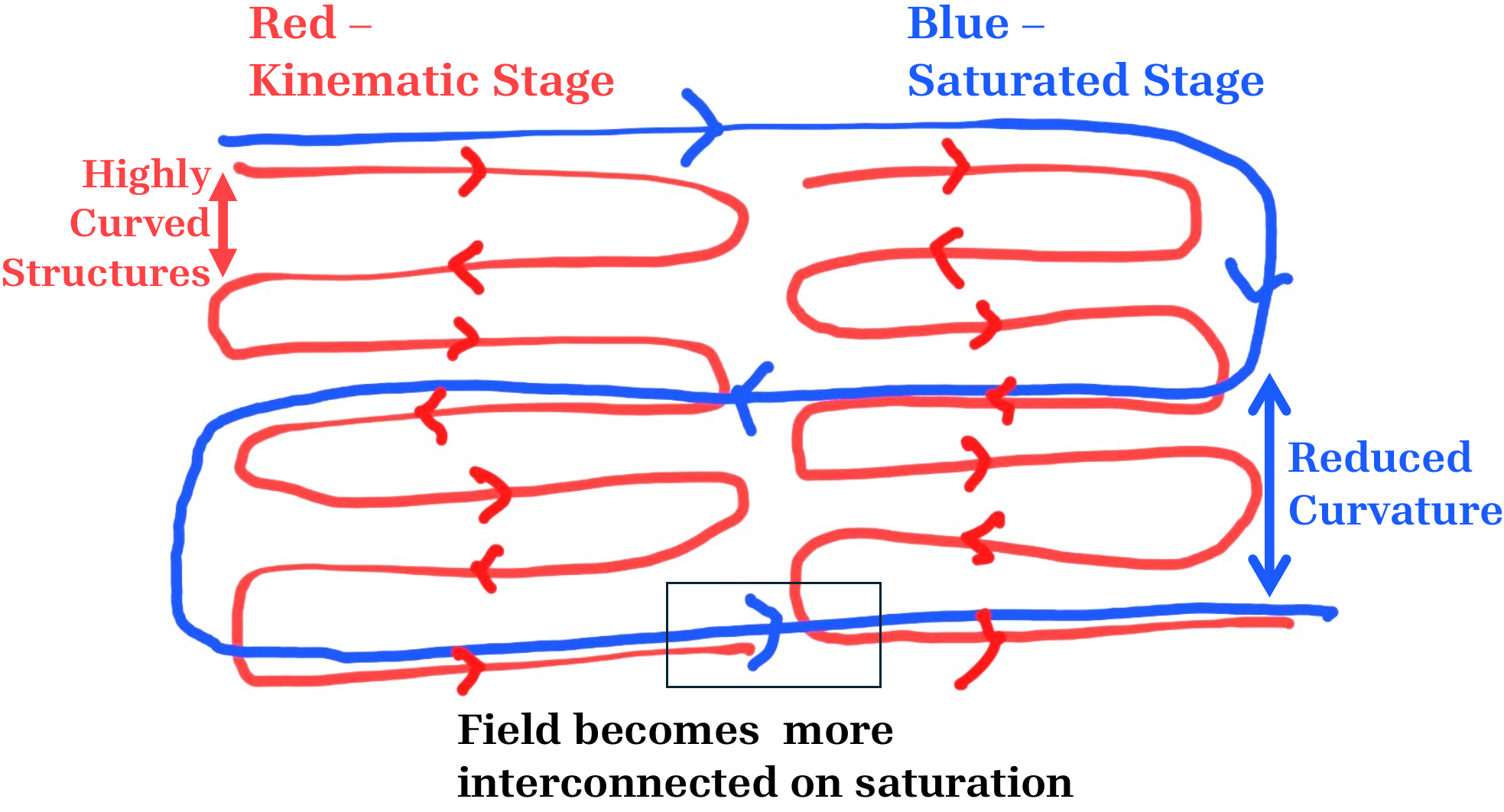}
		\caption{\revb{A schematic representation of the evolution of magnetic field structure from the kinematic to the saturated stage. Red lines represent the kinematic stage, characterised by tightly packed, highly curved field lines. Blue lines represent the saturated stage, in which the field lines interconnect and become less curved. Overall, this schematic illustrates that magnetic field structures become less curved and more interconnected upon saturation, consistent with our results showing statistically lower $V_2$ and more negative $V_3$ in the saturated stage (see \Fig{fig:MFs_2_3} and \Tab{tab:quant}).}}
		\label{fig:structure_changes}
	\end{figure}
	\revb{From our results (\Sec{sec:res}), we find that the magnetic structures become less curved (lower $V_2$) and more interconnected (more negative $V_3$) upon saturation. We illustrate how this can be physically interpreted in \Fig{fig:structure_changes}, especially for the strong-field regions (see \Fig{fig:MFs_2_3}). This can be interpreted as quantitative evidence for some key theoretical paradigms of the small-scale turbulent dynamo in the literature. 
		
		The relatively lower curvature of magnetic field structures in the saturated stage can be linked to the magnetic correlation length, which in turn reflects the influence of turbulent eddies. In the kinematic stage, small-scale eddies (of order the viscous scale) are most effective at amplifying magnetic fields \citep{Rincon2019}, and the magnetic field follows a Kazantsev spectrum \citep[with power at wavenumber $k$ scaling as $\propto k^{3/2}$,][]{Kazantsev1968}. This results in a magnetic correlation length that is significantly smaller than the turbulent driving scale. As the magnetic field grows and eventually saturates, larger scales also contribute; the magnetic power spectrum is modified, and the correlation length increases \citep{ChoR2009, BhatS2013, SurEA2018, SetaEA2020, seta_saturation_2021}. This implies that more ordered, larger-scale structures emerge as the magnetic field saturates. This picture is consistent with our finding of a decrease in the curvature of magnetic structures upon saturation.
		
		It has also been proposed that the efficiency of magnetic field line stretching, the mechanism responsible for magnetic field amplification \citep{Zeldovich1983}, statistically decreases at saturation \citep{Subramanian1998, SchekochihinEA2004, SetaEA2020}. Since stretching perpendicular to the magnetic field lines is most effective for amplification, a reduction in stretching efficiency is therefore expected to be accompanied by a decrease in the magnetic field curvature. This is consistent with our topological inference, which provides additional evidence of a statistical decline in the efficiency of field-line stretching.
		
		The reduction in curvature is also consistent with the ``folded-structure'' model of \citet{SchekochihinEA02, SchekochihinEA2004} for Prandtl number greater than or equal to unity. The model discusses how the magnetic fields are organised into folds, especially in the strong field regions, with direction reversals occurring at approximately the resistive scale (the fold-width) and the elongated segments are extended up to approximately the turbulent driving scale (the fold-length). Our result that the field becomes statistically less curved in the saturated stage is consistent with the prediction that these folds elongate as the dynamo saturates. The field organises into longer, more coherent structures (blue lines in \Fig{fig:structure_changes}), which have a lower average curvature compared to the highly curved, chaotic structures of the kinematic stage (red lines in \Fig{fig:structure_changes}).
		
		The increase in topological interconnectedness aligns with the mechanism proposed by \citet{blackman_overcoming_1996}, who argues that, for the large-scale magnetic dynamo to overcome the back-reaction from the small-scale fields in turbulent flows, it is helpful if the field organises into flux tubes capable of rapid reconnection. A highly interconnected small-scale field, as indicated by negative $V_3$, provides a physical regime that can support such rapid reconnection, also allowing the large-scale dynamo mechanism to operate despite the back-reaction from the small-scale dynamo-amplified fields \citep{blackman_overcoming_1996, Subramanian1998}.
		
		We note that these connections with the last two paradigms apply primarily to strong-field regions, where the field is organised into folded structures/flux tubes, whereas our results are derived for the entire field (see \Fig{fig:MFs}). Moreover, such folded structures may not be omnipresent \citep[e.g., see Figs.~5.5 and 5.6 in][]{BrandenburgS2005}. Although these analytical paradigms were derived for subsonic flows, they remain relevant as compressibility increases; however, the relative differences in $V_2$ and $V_3$ between the kinematic and saturated stages decrease with increasing compressibility (see \Fig{fig:MFs_2_3}). More analytical work is required to further develop these theoretical models and connect them to the detailed morphology of simulated non-Gaussian magnetic fields, especially in the compressible turbulent dynamo regime.
	}

	\section{Conclusions} \label{sec:con}
	Magnetic fields in protogalaxies are amplified to present-day strengths by a turbulent dynamo mechanism, which on a smaller $\lesssim 100\,\pc$ scale generates highly non-Gaussian magnetic fields. The weak initial seed field amplifies exponentially (kinematic stage) and then finally saturates to a statistically steady state (saturated stage) due to the back reaction from the strong, amplified magnetic fields. Using Minkowski functionals, we quantitatively characterise the morphology of such non-Gaussian magnetic fields obtained from numerical simulations of the turbulent dynamo, where the turbulence is driven with varying compressibility, which is characterised by the Mach number, $\mach = 0.1, 2, 5,$ and $10$ (representative of different phases of the interstellar medium within the galaxy).
	
	We confirm that the magnetic fields generated by the turbulent dynamo in both the kinematic and saturated stages at all $\mach$ are significantly different from their corresponding Gaussianised version with the same power spectra (\Sec{sec:GRFNGRF}). We conclusively demonstrate that the magnetic structures become relatively less curved and more interconnected as the turbulent dynamo saturates (\Sec{sec:kinsat}). We also show that the difference in magnetic morphology between the kinematic and saturated stages statistically decreases with increasing Mach number. This is further explained by the interplay between the density fluctuations in turbulence and the role of back reaction in the turbulent dynamo (\revb{\Sec{sec:disa}), which} provides a deeper understanding of the turbulent dynamo saturation mechanism. \revb{Finally, we discuss our findings in the context of existing analytical models in the literature (\Sec{sec:disb}) to highlight the analysis as a promising tool for comparison with small-scale turbulent dynamo theories.}

	\revb{Also, Minkowski functionals provide a rich avenue for probing the higher-order statistical properties of magnetic fields, features that encapsulate rich morphological information but are absent in standard two-point statistics. As recently shown by \cite{DuttaEA2024}, they are powerful tools for deconstructing the relationship between the morphology of polarisation observations and the underlying turbulent driving scale of astrophysical plasma. Polarisation observables are inherently sensitive to the field's vector properties, and thus it is important to understand the 3D magnetic morphology to interpret projected 2D observations and related possible degeneracies. We suggest that further careful study and the standardisation of Minkowski functionals for the 3D to 2D projections \citep[also, see complementary efforts in][]{MakarenkoFS2015} will enable in-depth studies of non-Gaussian magnetic fields in astrophysical observations}. \rev{In particular, \revb{this} will be useful for rich and dense radio polarisation data from current (e.g.~Australian SKA Pathfinder surveys: SPICE-RACS; \citealt{ThomsonEA2023}, POSSUM; \citealt{GaenslerEA2025} and MPIfR-MeerKAT Galactic Plane Survey; \citealt{PadmanabhEA2023}) and upcoming (SKA; \citealt{HealdEA2020}) facilities.
	}
	
	\section*{Acknowledgements}
	\revb{We thank the anonymous referee and Eric Blackman for their very useful comments and suggestions.} SMB thanks Eswaraiah Chakali for his constant support and mentorship throughout the project. We also thank Eswaraiah Chakali and Jessy Jose for providing access to computational resources in their respective laboratories at IISER Tirupati. We are grateful to Thomas Buchert for generously sharing his code for evaluating Minkowski Functionals \citep{schmalzing_beyond_1997}, which our Python implementation draws on significantly. AS is supported by the Australian Research Council through the Discovery Early Career Researcher Award (DECRA) Fellowship (project~DE250100003) funded by the Australian Government and the Australia-Germany Joint Research Cooperation Scheme of Universities Australia (UA--DAAD, 2025--2026).
	
	\section*{Data Availability}
	The simulated data and derived results will be shared on a reasonable request to the authors.

	
	
	\bibliographystyle{mnras}
	\bibliography{master_references}


	\bsp	
	\label{lastpage}
\end{document}